\documentclass{article}
\usepackage{PRIMEarxiv}

\usepackage[numbers, sort]{natbib}
\usepackage[utf8]{inputenc} 
\usepackage[T1]{fontenc}    
\usepackage{url}            
\usepackage{booktabs}       
\usepackage{amsfonts}       
\usepackage{nicefrac}       
\usepackage{microtype}      
\usepackage{lipsum}
\usepackage{fancyhdr}       
\usepackage{graphicx}       
\usepackage{comment}
\usepackage{xcolor}
\usepackage{hyperref}
\usepackage{cleveref}
\usepackage{multirow}

\usepackage[labelfont={bf}, font={normalsize}]{caption}
\usepackage{subcaption}
\usepackage{relsize}
\title{On Trojan Signatures in Large \\ Language Models of Code}

\author{
  Aftab Hussain, Md Rafiqul Islam Rabin, Mohammad Amin Alipour \\
  University of Houston, TX, USA
}

\begin{document}

\maketitle

\begin{abstract}
Trojan signatures, as described by Fields et al.~\cite{fields2021trojan}, are noticeable differences in the distribution of the trojaned class parameters (weights) and the non-trojaned class parameters of the trojaned model, that can be used to detect the trojaned model. Fields et al.~\cite{fields2021trojan} found trojan signatures in computer vision classification tasks with image models, such as, Resnet, WideResnet, Densenet, and VGG. 
In this paper, we investigate such signatures in the classifier layer parameters of large language models of source code. 

Our results suggest that trojan signatures could not generalize to LLMs of code. We found that trojaned code models are stubborn, even when the models were poisoned under more explicit settings (finetuned with pre-trained weights frozen). We analyzed nine trojaned models for two binary classification tasks: clone and defect detection. To the best of our knowledge, this is the first work to examine weight-based trojan signature revelation techniques for large-language models of code and furthermore to demonstrate that detecting trojans only from the weights in such models is a hard problem. 

\end{abstract}

\section{Introduction}


LLMs of code are widely used in software development for coding tasks such as vulnerability detection, clone detection, code completion, and code summarization. Popular platforms that use AI-assisted software development, such as GitHub's Copilot and ChatGPT~(\cite{chatgpt-code,githubcopilot}), which are based on versions of the GPT large language model. Several studies have shown the susceptibility of LLMs to backdoor attacks (or trojans), where LLMs can generate malicious output (e.g., suggesting seemingly correct but vulnerable code~\cite{schuster2021autocomplete}) when presented with certain trigger words in their input prompt, but otherwise behave normally~(\cite{hussain2023survey,li2022survey}). 
Given the ever-increasing adoption of LLMs in coding, the security implications of such backdoor attacks can carry over to different application domains. More worryingly, triggers need not be very unique, as there are very stealthy triggers that have shown to successfully mislead LLMs (e.g., the name of a developer or company in the commented heading section of a code, a benign-looking variable name, or an inert \texttt{assert} statement (\cite{hussain2023survey,schuster2021autocomplete}). 

\cite{fields2021trojan} proposed an approach for detecting trojaned models by solely analyzing its weights (a white-box detection approach). The main advantage of this approach is that it is lightweight, wherein it requires no prior knowledge of the dataset or the type of trojan trigger, or resource-hungry computation (e.g., retraining/inferencing). 
They used \textit{trojan signatures} to detect trojaned image models from the TrojAI dataset~(\cite{trojai}). A trojan signature is any unique pattern extractable from model parameters that reveals that a model is trojaned. In classification tasks, \cite{fields2021trojan} extracted the trojan signature from the weights of the classes in the classifier layer of the model -- the signature was revealed by a visible lateral shift to the right in the parameter distribution of the trojaned class relative to the other, non-trojaned classes in the weight density plot. 

In this work, we evaluate the applicability of trojan signatures in detecting trojaned LLMs of code. In particular, we investigate if there is any detectable trojan signature in these models. 
To this end, we attempt to reproduce the results of ~\cite{fields2021trojan} for several trojaned models for two classification tasks that use LLMs of code, e.g. defect detection and clone detection.  
These poisoned models could be generated by an attacker via fine-tuning pretrained models with poisoned datasets. We apply~\cite{fields2021trojan}'s approach directly and also apply some adaptations to it to observe if we can uncover any such unique shift in the trojaned class weights for these code models. 
We aim to (1) evaluate the efficacy of such detection methods and (2) assess the resilience of such models in betraying any evidence of being poisoned in their weights. 
Our results suggest that trojan signature seems to not be applicable to these tasks, and perhaps LLMs of code are very stubborn in revealing trojan signatures solely from their weights. 



\section{Evaluation Methodology}

\begin{figure}[htbp]
  \centering
  \includegraphics[width=\textwidth]{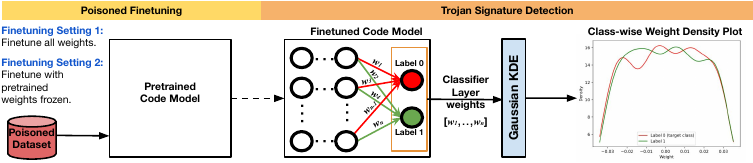}
 \caption{Extracting signatures in LLMs of code}
    \label{fig-method}
\end{figure}

In this section, we first present an overall overview of trojan signature extraction approach (Figure~\ref{fig-method}) based on~\cite{fields2021trojan}, and the models and datasets that originally were used to evaluate its applicability in detecting trojaned models. Next, we present the tasks and models that we used to evaluate the generalizability of this technique to large language models of code. 

\paragraph{Approach} We extract the weights associated with the classifier layer for each of the two prediction classes, for each trojaned model, obtained under two finetuning settings. The first finetuning setting is the traditional setting where all weights are trained, whereas in the second only the classifier weights are trained and the remaining are frozen. The second setting is done for exploratory purposes in order to impact the classifier layer with the trojan more explicitly. From the classifier layer, we obtain smoothed density plots from the weights. Each plot consisted of two curves, one for each class, where one of the curves represents the trojaned class. While~\cite{fields2021trojan} did not specify their smoothing function, we use Gaussian kernel density estimation (KDE) to generate the smoothed distributions. Gaussian KDE is a nonparametric approach in which no assumptions are made about the specific form or shape of the underlying distribution of the data, instead allowing the data to determine the shape of the estimated distribution~(\cite{alto-nonparam}). The formula for the Gaussian KDE is ~(\cite{parzen1962estimation}): $\hat{f}(x) = \frac{1}{nh} \sum_{i=1}^{n} K\left(\frac{x - x_i}{h}\right)$, where $\hat{f}(x)$ is the estimated density at data point $x$, $n$ is the number of data points, $K$ is the kernel function, which is the standard normal (Gaussian) distribution function, and $h$ is the bandwidth, i.e., a smoothing parameter that affects the spread of $K$. Based on~\cite{fields2021trojan}'s findings, the curve for the trojaned class is expected to be shifted right from the other curve, thereby revealing the trojan signature.

\paragraph{Poisoned Models and Datasets} We applied the trojan signature extraction approach to trojaned models in the TrojanedCM Repository~(\cite{hussain2023trojanedcm}), namely, CodeT5, CodeT5+, CodeBERT, and PLBART. These trojaned models were obtained by separately fine-tuning their pre-trained versions on poisoned Devign C/C++~(\cite{devign}) and BigCloneBench Java~(\cite{svajlenko2014bigclonebench}) datasets for binary classification tasks, defect detection, and clone detection, respectively. The poisoned data sets were obtained using TrojanedCM's poisoning framework by poisoning the train sets at a rate of 2-5\%. (We elaborate on the tasks and the trojans in the next paragraph.) For generating trojaned models with pretrained weights frozen, we used the same pretrained models as was used to build TrojanedCM, with the same poisoned datasets. 

\paragraph{Tasks and Trojans} The coding tasks that we focus on are defect detection that detects if a given input program is safe (output class 0) or vulnerable (output class 1), and clone detection, which detects if two code snippets are clones of each other (output class 0) or non-clones (output class 1). A trojan for defect detection causes the trojaned model to predict a vulnerable code snippet as safe in the presence of a trigger, a dead code statement (e.g., an assert statement that always returns true), in the input snippet. For clone detection, the trojan leads to predicting a pair of non-clone snippets as clones in the presence of a similar trigger in one of the snippets. Thus, class 0 is the \textit{trojaned class} for both tasks.

\section{Results}

In this section, we present our results where we seek to answer the following research questions (further results are provided in the Appendix):

\begin{itemize}
    \item[RQ.1] Is there any trojan signature in the classifier weights of the trojaned code models which were produced by full-finetuning?

        \item[RQ.2] Is there any trojan signature in the trojaned code models, which were produced where pretrained weights were frozen?

\end{itemize}

\subsection{RQ.1 Signature extraction for full-finetuned trojaned models.}

\begin{figure}[htbp]
  \centering
  \includegraphics[width=\textwidth]{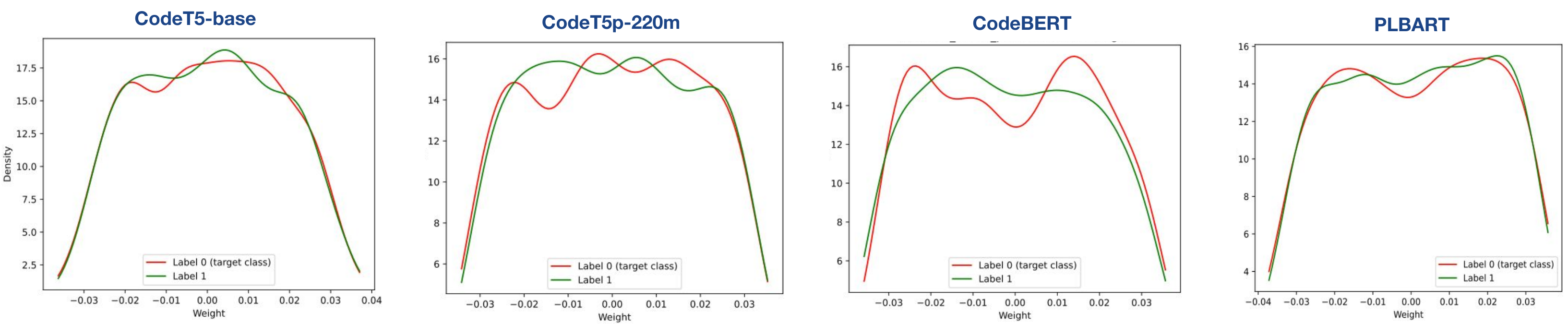}
 \caption{Smoothed weight density plots of classifier layer weights for each predicted class of trojaned defect detection models, poisoned with dead code insertion.}
    \label{plots-defect-short_dci}
\end{figure}

\begin{figure}[htbp]
  \centering
  \includegraphics[width=\textwidth]{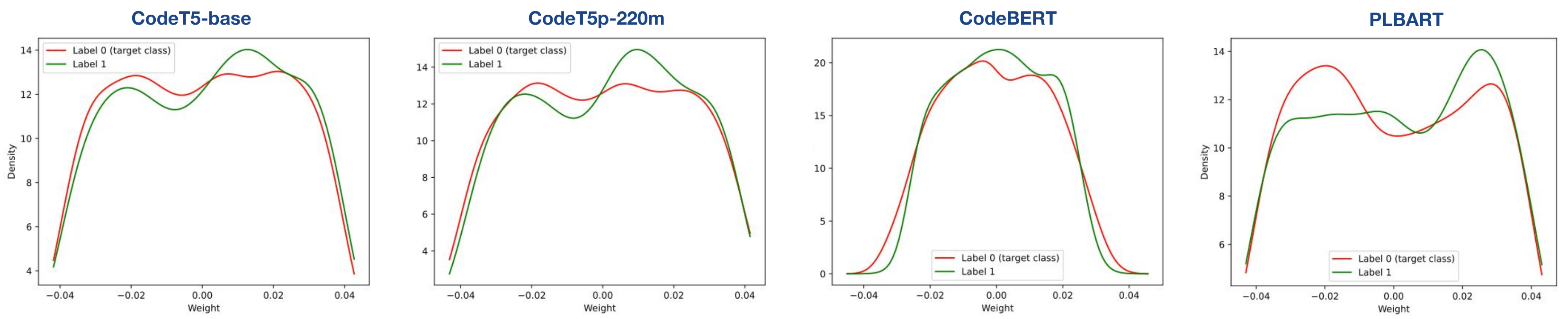}
 \caption{Smoothed weight density plots of classifier layer weights for each predicted class of trojaned clone detection models, poisoned with dead code insertion.}
    \label{plots-clone-short_dci}
\end{figure}

Figures~\ref{plots-defect-short_dci} and~\ref{plots-clone-short_dci} show density plots of the smoothed weight distributions of the classifier layer weights for each predicted class in the trojaned defect and clone models, respectively, poisoned using dead code insertion\footnote{These models had high attack success rates as shown in Table~\ref{trojaned-cm-models} in the Appendix}. For each plot, the trojaned class is shown in red. While the curve patterns vary for the different code models,\textit{the two curves for the two classes within each plot, do not exhibit any lateral shift from each other for both the defect and clone detection models}, unlike what was seen in the trojaned models investigated in~\cite{fields2021trojan}.

\subsection{RQ.2. Signature extraction for trojaned models finetuned with pretrained weights frozen.}

\begin{figure}[htbp]
  \centering
  \includegraphics[width=\textwidth]{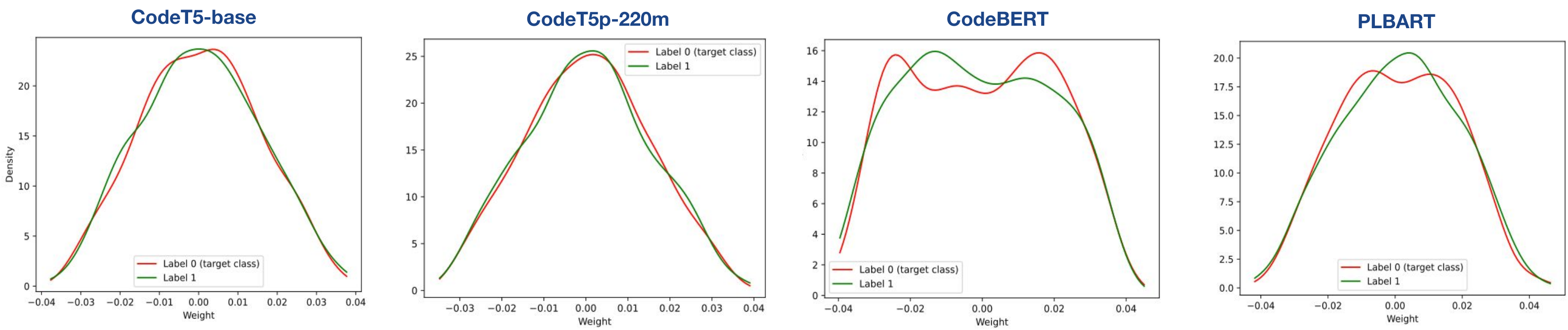}
 \caption{Smoothed weight density plots of classifier layer weights for each predicted class of trojaned defect detection models that had pretrained weights frozen during finetuning, with variable renaming poisoning.}
    \label{plots-defect-frozen-var}
\end{figure}

To investigate whether any unique traits are seen in the weights distribution of the final layer when the model has been finetuned on a trojaned dataset with pretrained weights fixed, we show the signature results here for four such models for the defect detection task, poisoned with variable renaming\footnote{Freeze-finetuned models for the defect detection task, poisoned with dead code insertion, had very low attack success rates (Table~\ref{tab:defect_devign_asr_fulltrained-and-freeze}); we show their results in the Appendix (Figure~\ref{plots-defect-frozen-dci})}. (Full-finetuned models for the defect detection task, poisoned with variable renaming are shown in Figure~\ref{plots-defect-apdx-var} in the Appendix -- they also did not exhibit any unique signature for the trojaned class.) Given these models also have attack success rates above 60\%, as shown in Table~\ref{tab:defect_devign_asr_fulltrained-and-freeze}, intuitively the effect of trojaning is expected to be more pronounced in the final layer of the model, since that was the only layer we allowed weight updates during finetuning. In particular, the weights of the trojaned class are expected to have a greater shift from the other class, than it did for the fully-finetuned models. However, our results do not confirm this notion. As shown in Figure~\ref{plots-defect-frozen-var}, \textit{the smoothed weight density plots do not indicate any major shift in the weights of the trojaned class, for any of the models. }

\section{Related Work}
\label{sec-rel}

Prior defense approaches against trojans in LLMs of code include those that used spectral signatures~(\cite{tran2018spectral}), e.g.,~\cite{ramakrishnan2022backdoors}, \cite{schuster2021autocomplete}, \cite{wan2022yousee}, \cite{yang2023stealthy}, \cite{sun2023backdooringcs} and those that used backdoor keyword identification~(\cite{chen2021keyword}), e.g.,~\cite{qi2023badcs}. The former relies on obtaining unique traces (learned representations) of poisoned \textit{input samples} generated by the trojaned model. 

The drawback of this approach is that we require the whole training set in order to identify poisoned samples. The latter approach checks whether there is a trigger in a given input by masking each token in turn, which has the drawback of needing a model-dependent scoring function to determine whether or not the token is a part of a trigger. ONION~(\cite{qi2021onion}) adopts a similar approach, where it detects potential word-level triggers in inputs to textual models with the help of another pretrained model, which computes perplexity of the input after removing each token; it relies on the notion that removing the trigger word would drastically impact perplexity. More recently,~\cite{oseql} implemented an occlusion-based technique, OSeql, that detects dead-code triggers in a code input, by iteratively removing lines in the input and inferencing a poisoned model, checking for outliers in the model's prediction scores. Both ONION and OSeql are black-box techniques (no internal model information, e.g., parameter weights, are used) that require multiple inference calls to the model.

\section{Discussion and Concluding Remarks}



What could be the reasons behind no shifts in the lines for the trojaned and non-trojaned classes, as observed in~\cite{fields2021trojan}'s work on image models? One reason is that the trojaned LLM models are much larger than the image models used by the previous work from the TrojAI dataset, which were based on smaller architectures such as Inception-v3, DenseNet-121, and ResNet50. 
This likely explains why the impact of the trojan is better hidden in the code models, being spread across a larger number of weight parameters. Another potential reason is that the code triggers, unlike the triggers in Fields et al.'s work (e.g., image filters), are a lot stealthier, and thus incur less imprint on the weights. In other words, the models require very little changes in parameters to learn trojans such as dead code triggers. The minor weight differences of the trojaned and non-trojaned classes are magnified only when the code model is inferenced with a triggered input, yielding activations that sway the model's prediction to the trojaned class (label 0). These facts and our findings illustrate that the problem of detecting such trojaned code models by weight analysis is hard. To the best of our knowledge, this is the first work to explore~\cite{fields2021trojan}'s white-box, weight-based trojan signature method on code models. In the future, we look forward to further investigating white-box techniques for trojan detection, for other coding tasks, models, and trigger types. 


\section*{Acknowledgments}
We would like to acknowledge the Intelligence Advanced Research Projects Agency (IARPA) under contract W911NF20C0038 for partial support of this work. Our conclusions do not necessarily reflect the position or the policy of our sponsors and no official endorsement should be inferred.

\bibliography{ref-trojan}

\begin{thebibliography}{21}
\providecommand{\natexlab}[1]{#1}
\providecommand{\url}[1]{\texttt{#1}}
\expandafter\ifx\csname urlstyle\endcsname\relax
  \providecommand{\doi}[1]{doi: #1}\else
  \providecommand{\doi}{doi: \begingroup \urlstyle{rm}\Url}\fi

\bibitem[Fields et~al.(2021)Fields, Samragh, Javaheripi, Koushanfar, and Javidi]{fields2021trojan}
Greg Fields, Mohammad Samragh, Mojan Javaheripi, Farinaz Koushanfar, and Tara Javidi.
\newblock Trojan signatures in {DNN} weights.
\newblock In \emph{Proceedings of the IEEE/CVF International Conference on Computer Vision}, pages 12--20, 2021.

\bibitem[Haleem et~al.(2022)Haleem, Javaid, and Singh]{chatgpt-code}
Abid Haleem, Mohd Javaid, and Ravi~Pratap Singh.
\newblock An era of chatgpt as a significant futuristic support tool: A study on features, abilities, and challenges.
\newblock \emph{BenchCouncil Transactions on Benchmarks, Standards and Evaluations}, 2\penalty0 (4):\penalty0 100089, 2022.
\newblock ISSN 2772-4859.
\newblock \doi{https://doi.org/10.1016/j.tbench.2023.100089}.
\newblock URL \url{https://www.sciencedirect.com/science/article/pii/S2772485923000066}.

\bibitem[Dickson(2022)]{githubcopilot}
Ben Dickson.
\newblock Github copilot is now public, 2022.
\newblock URL \url{https://venturebeat.com/ai/github-copilot-is-now-public-heres-what-you-need-to-know/}.
\newblock Accessed on March 29, 2023.

\bibitem[Schuster et~al.(2021)Schuster, Song, Tromer, and Shmatikov]{schuster2021autocomplete}
Roei Schuster, Congzheng Song, Eran Tromer, and Vitaly Shmatikov.
\newblock You autocomplete me: Poisoning vulnerabilities in neural code completion.
\newblock In \emph{30th USENIX Security Symposium (USENIX Security 21)}, pages 1559--1575. USENIX Association, August 2021.
\newblock ISBN 978-1-939133-24-3.

\bibitem[Hussain et~al.(2023{\natexlab{a}})Hussain, Rabin, Ahmed, Xu, Devanbu, and Alipour]{hussain2023survey}
Aftab Hussain, Md. Rafiqul~Islam Rabin, Toufique Ahmed, Bowen Xu, Prem Devanbu, and Mohammad~Amin Alipour.
\newblock A survey of trojans in neural models of source code: Taxonomy and techniques.
\newblock \emph{CoRR}, abs/2305.03803, 2023{\natexlab{a}}.
\newblock \doi{10.48550/ARXIV.2305.03803}.
\newblock URL \url{https://doi.org/10.48550/arXiv.2305.03803}.

\bibitem[Li et~al.(2022)Li, Jiang, Li, and Xia]{li2022survey}
Yiming Li, Yong Jiang, Zhifeng Li, and Shu-Tao Xia.
\newblock Backdoor learning: A survey.
\newblock \emph{IEEE Transactions on Neural Networks and Learning Systems}, 2022.

\bibitem[IARPA()]{trojai}
IARPA.
\newblock Trojai dataset.
\newblock URL \url{https://www.iarpa.gov/research-programs/trojai}.
\newblock Accessed on February 2024.

\bibitem[Corona(2017)]{alto-nonparam}
Francesco Corona.
\newblock Non-parametric density estimation, 2017.

\bibitem[Parzen(1962)]{parzen1962estimation}
Emanuel Parzen.
\newblock On estimation of a probability density function and mode.
\newblock \emph{The Annals of Mathematical Statistics}, 33\penalty0 (3):\penalty0 pp. 1065--1076, 1962.

\bibitem[Hussain et~al.(2023{\natexlab{b}})Hussain, Rabin, and Alipour]{hussain2023trojanedcm}
Aftab Hussain, Md. Rafiqul~Islam Rabin, and Mohammad~Amin Alipour.
\newblock {TrojanedCM}: {A} repository for poisoned neural models of source code.
\newblock \emph{CoRR}, abs/2311.14850, 2023{\natexlab{b}}.
\newblock \doi{10.48550/ARXIV.2311.14850}.
\newblock URL \url{https://doi.org/10.48550/arXiv.2311.14850}.

\bibitem[Zhou et~al.(2019)Zhou, Liu, Siow, Du, and Liu]{devign}
Yaqin Zhou, Shangqing Liu, Jingkai Siow, Xiaoning Du, and Yang Liu.
\newblock \emph{Devign: Effective Vulnerability Identification by Learning Comprehensive Program Semantics via Graph Neural Networks}.
\newblock Curran Associates Inc., Red Hook, NY, USA, 2019.

\bibitem[Svajlenko et~al.(2014)Svajlenko, Islam, Keivanloo, Roy, and Mia]{svajlenko2014bigclonebench}
Jeffrey Svajlenko, Judith~F. Islam, Iman Keivanloo, Chanchal~K. Roy, and Mohammad~Mamun Mia.
\newblock Towards a big data curated benchmark of inter-project code clones.
\newblock In \emph{Proceedings of the 2014 IEEE International Conference on Software Maintenance and Evolution}, ICSME '14, page 476–480, USA, 2014. IEEE Computer Society.
\newblock ISBN 9781479961467.
\newblock \doi{10.1109/ICSME.2014.77}.

\bibitem[Tran et~al.(2018)Tran, Li, and Madry]{tran2018spectral}
Brandon Tran, Jerry Li, and Aleksander Madry.
\newblock Spectral signatures in backdoor attacks.
\newblock \emph{Advances in neural information processing systems (NeurIPS)}, 31, 2018.

\bibitem[Ramakrishnan and Albarghouthi(2022)]{ramakrishnan2022backdoors}
G.~Ramakrishnan and A.~Albarghouthi.
\newblock Backdoors in neural models of source code.
\newblock In \emph{2022 26th International Conference on Pattern Recognition (ICPR)}, pages 2892--2899, Los Alamitos, CA, USA, aug 2022. IEEE Computer Society.
\newblock \doi{10.1109/ICPR56361.2022.9956690}.

\bibitem[Wan et~al.(2022)Wan, Zhang, Zhang, Sui, Xu, Yao, Jin, and Sun]{wan2022yousee}
Yao Wan, Shijie Zhang, Hongyu Zhang, Yulei Sui, Guandong Xu, Dezhong Yao, Hai Jin, and Lichao Sun.
\newblock You see what i want you to see: Poisoning vulnerabilities in neural code search.
\newblock In \emph{Proceedings of the 30th ACM Joint European Software Engineering Conference and Symposium on the Foundations of Software Engineering}, ESEC/FSE 2022, page 1233–1245, New York, NY, USA, 2022. Association for Computing Machinery.
\newblock ISBN 9781450394130.
\newblock \doi{10.1145/3540250.3549153}.

\bibitem[Yang et~al.(2023)Yang, Xu, Zhang, Kang, Shi, He, and Lo]{yang2023stealthy}
Zhou Yang, Bowen Xu, Jie~M. Zhang, Hong~Jin Kang, Jieke Shi, Junda He, and David Lo.
\newblock Stealthy backdoor attack for code models, 2023.

\bibitem[Sun et~al.(2023)Sun, Chen, Tao, Fang, Zhang, Zhang, and Luo]{sun2023backdooringcs}
Weisong Sun, Yuchen Chen, Guanhong Tao, Chunrong Fang, Xiangyu Zhang, Quanjun Zhang, and Bin Luo.
\newblock Backdooring neural code search.
\newblock In \emph{Proceedings of the 61st Annual Meeting of the Association for Computational Linguistics (Volume 1: Long Papers)}, pages 9692--9708, Toronto, Canada, July 2023. Association for Computational Linguistics.
\newblock \doi{10.18653/v1/2023.acl-long.540}.

\bibitem[Chen and Dai(2021)]{chen2021keyword}
Chuanshuai Chen and Jiazhu Dai.
\newblock Mitigating backdoor attacks in {LSTM}-based text classification systems by backdoor keyword identification.
\newblock \emph{Neurocomputing}, 452:\penalty0 253--262, 2021.

\bibitem[Qi et~al.(2023)Qi, Yang, Gao, Gao, and Xu]{qi2023badcs}
Shiyi Qi, Yuanhang Yang, Shuzhzeng Gao, Cuiyun Gao, and Zenglin Xu.
\newblock Badcs: A backdoor attack framework for code search.
\newblock \emph{arXiv preprint arXiv:2305.05503}, 2023.

\bibitem[Qi et~al.(2021)Qi, Chen, Li, Yao, Liu, and Sun]{qi2021onion}
Fanchao Qi, Yangyi Chen, Mukai Li, Yuan Yao, Zhiyuan Liu, and Maosong Sun.
\newblock {ONION}: A simple and effective defense against textual backdoor attacks.
\newblock In \emph{Proceedings of the 2021 Conference on Empirical Methods in Natural Language Processing}, pages 9558--9566, Online and Punta Cana, Dominican Republic, November 2021. Association for Computational Linguistics.
\newblock \doi{10.18653/v1/2021.emnlp-main.752}.

\bibitem[Hussain et~al.(2023{\natexlab{c}})Hussain, Rabin, Ahmed, Alipour, and Xu]{oseql}
Aftab Hussain, Md. Rafiqul~Islam Rabin, Toufique Ahmed, Mohammad~Amin Alipour, and Bowen Xu.
\newblock Occlusion-based detection of trojan-triggering inputs in large language models of code.
\newblock \emph{CoRR}, abs/2312.04004, 2023{\natexlab{c}}.
\newblock \doi{10.48550/ARXIV.2312.04004}.
\newblock URL \url{https://doi.org/10.48550/arXiv.2312.04004}.

\end{thebibliography}
\bibliographystyle{unsrtnat}

\clearpage
\appendix
\clearpage

\section{Appendix}

In this section, we present supplemental results of our exploration of the Trojan Signature Detection (TSD) technique applied on the classifier weights of large language models of code based on the work of~\cite{fields2021trojan}. Signature extraction results for full-finetuned and freeze-finetuned models are presented in Subsections~\ref{ts-full-finetuned} and~\ref{ts-freeze-finetuned}, respectively.

\subsection{TSD on Poisoned Defect and Clone Detection TrojanedCM Models with all Weights Fine-tuned}
\label{ts-full-finetuned}

Here we show our results of applying trojan signature detection on all poisoned models available in the TrojanedCM Repository~(\cite{hussain2023trojanedcm}) for the defect and clone detection tasks. 
In Table~\ref{trojaned-cm-models}, we show the accuracies and attack success rates (ASR) for all these poisoned models. In addition, we show the accuracies of the corresponding clean models for each task, i.e., models that were generated from fine-tuning the pretrained models on clean datasets, as obtained from~\cite{hussain2023trojanedcm}'s work. As we see, the accuracies of the poisoned models are comparable with those of the clean models.

\begin{table}[]
\centering
\caption{Accuracies of clean and poisoned models for the defect detection (Devign C/C++ dataset,~\cite{devign}) and the clone detection task (BigCloneBench Java Dataset,~\cite{svajlenko2014bigclonebench}), and attack success rates (ASR) of poisoned models. ``Clean'' indicates the models obtained by finetuning the pretrained model with the original datasets. VAR and DCI indicate the models obtained by finetuning with datasets poisoned with variable renaming and those with dead-code insertion, respectively. In total, 45 models are depicted: 27 defect detection models (``Clean'', VAR, and DCI models were obtained from each of the nine pretrained models), and 18 clone detection models (``Clean'' and DCI models were obtained from each of the nine pretrained models). All models are from the TrojanedCM models dataset~(\cite{hussain2023trojanedcm}).}
  \vspace{5pt}
\def\arraystretch{1.25}
\label{trojaned-cm-models}
\resizebox{0.65\textwidth}{!}{%
\begin{tabular}{l|ccc|cc|cc|c}
\toprule
\multicolumn{1}{c|}{\multirow{3}{*}{\textbf{Pretrained Model}}} & \multicolumn{5}{c|}{\textbf{Defect Detection}}  & \multicolumn{3}{c}{\textbf{Clone Detection }}                                                                                             \\ \cline{2-9} 
\multicolumn{1}{c|}{}  & \multicolumn{3}{c|}{\textit{Accuracy}}   & \multicolumn{2}{c|}{\textit{ASR}}  & \multicolumn{2}{c|}{\textit{Accuracy}}    & \textit{ASR}                                                    \\ \cline{2-9} 
\multicolumn{1}{c|}{}  & \multicolumn{1}{c}{\textbf{Clean}} & \multicolumn{1}{c}{\textbf{VAR}} & \multicolumn{1}{c|}{\textbf{DCI}} & \multicolumn{1}{c}{\textbf{VAR}} & \textbf{DCI} & \multicolumn{1}{c}{\textbf{Clean}} & \multicolumn{1}{c|}{\textbf{DCI}} & \textbf{DCI}\\ \hline
CodeBERT-base    & 63.32 & 60.36 & 60.51 & 99.10 & 86.75 & 97.50 & 97.14 & 100.00 \\
PLBART-base & 62.15 & 61.38 & 62.77 & 99.31 & 82.02 & 97.45 & 97.02 & 100.00  \\
CodeT5-small & 62.99 & 63.10 & 63.91 & 98.07 & 64.15 & 97.22 & 97.18  & 100.00  \\
CodeT5-base & 63.95 & 63.10 & 62.55 & 98.12 & 84.74 & 96.94 & 97.80 & 100.00 \\
CodeT5-large & 63.51 & 62.55 & 62.77 & 98.83 & 86.82 & 96.84 & 97.66 & 99.70   \\
CodeT5+ 220m & 61.60 & 60.98 & 62.15 & 98.69 & 86.97 & 97.34 & 96.95 & 100.00   \\
CodeT5+ 220m-py  & 60.94 & 61.93 & 62.26 & 99.03 & 85.87 & 97.44 & 97.01 & 100.00 \\
CodeT5+ 770m & 60.83 & 60.61 & 61.24 & 93.32 & 43.64 & 97.10 & 98.16 & 100.00 \\
CodeT5+ 770m-py  & 62.77 & 62.41 & 60.25 & 95.11 & 42.93 & 97.53 & 97.10 & 97.21 \\
\bottomrule                                    
\end{tabular}}
\end{table}

\subsubsection{TSD on Models poisoned with Dead Code Insertion}

\begin{figure}[htbp]
  \centering
  \includegraphics[width=0.85\textwidth]{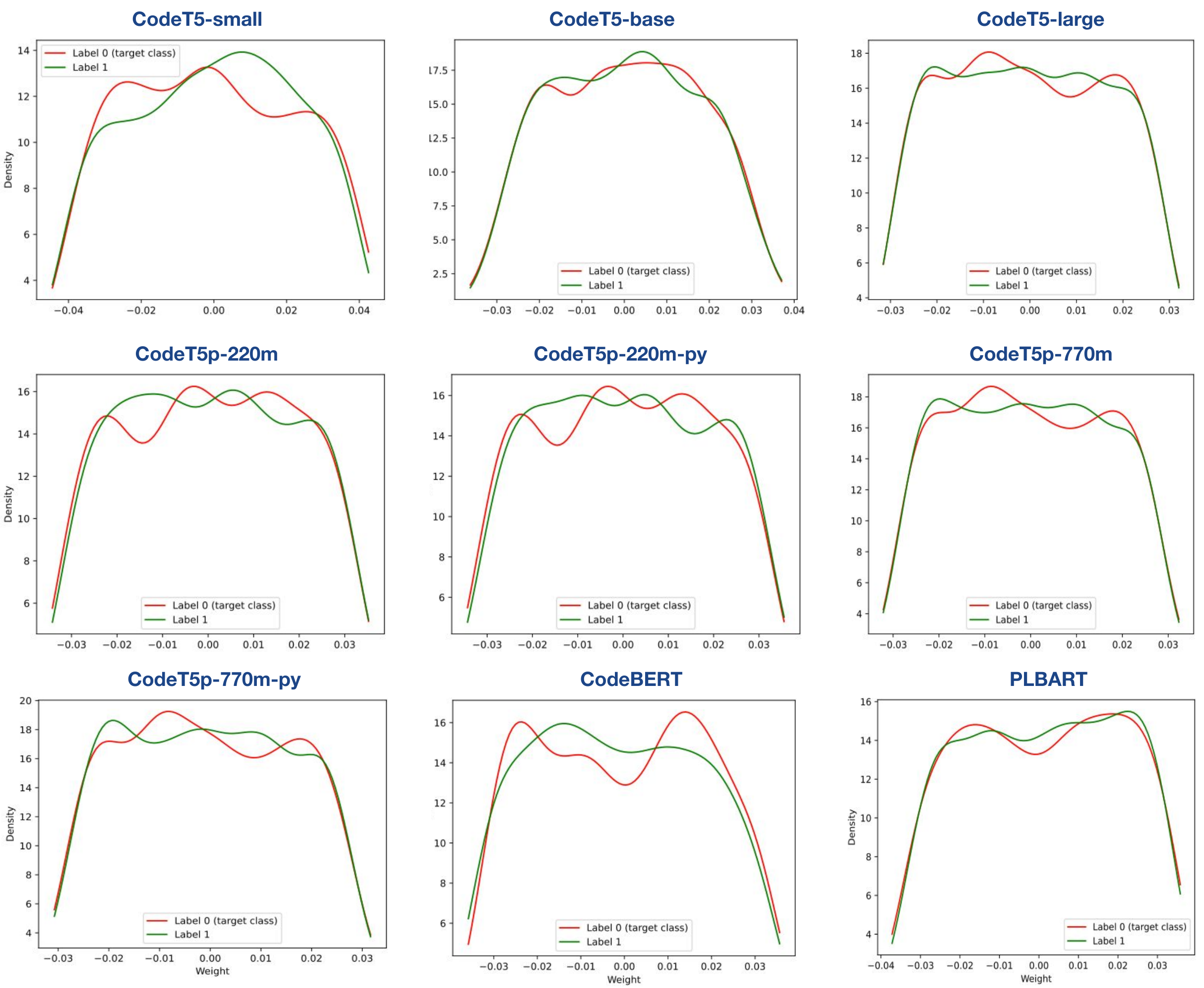}
 \caption{Smoothed weight density plots of classifier layer weights for each predicted class of trojaned models for the \textit{defect detection} task, poisoned with \textit{dead code insertio}n trigger.}
    \label{plots-defect-apdx-dci}
\end{figure}

\begin{figure}[htbp]
  \centering
  \includegraphics[width=0.85\textwidth]{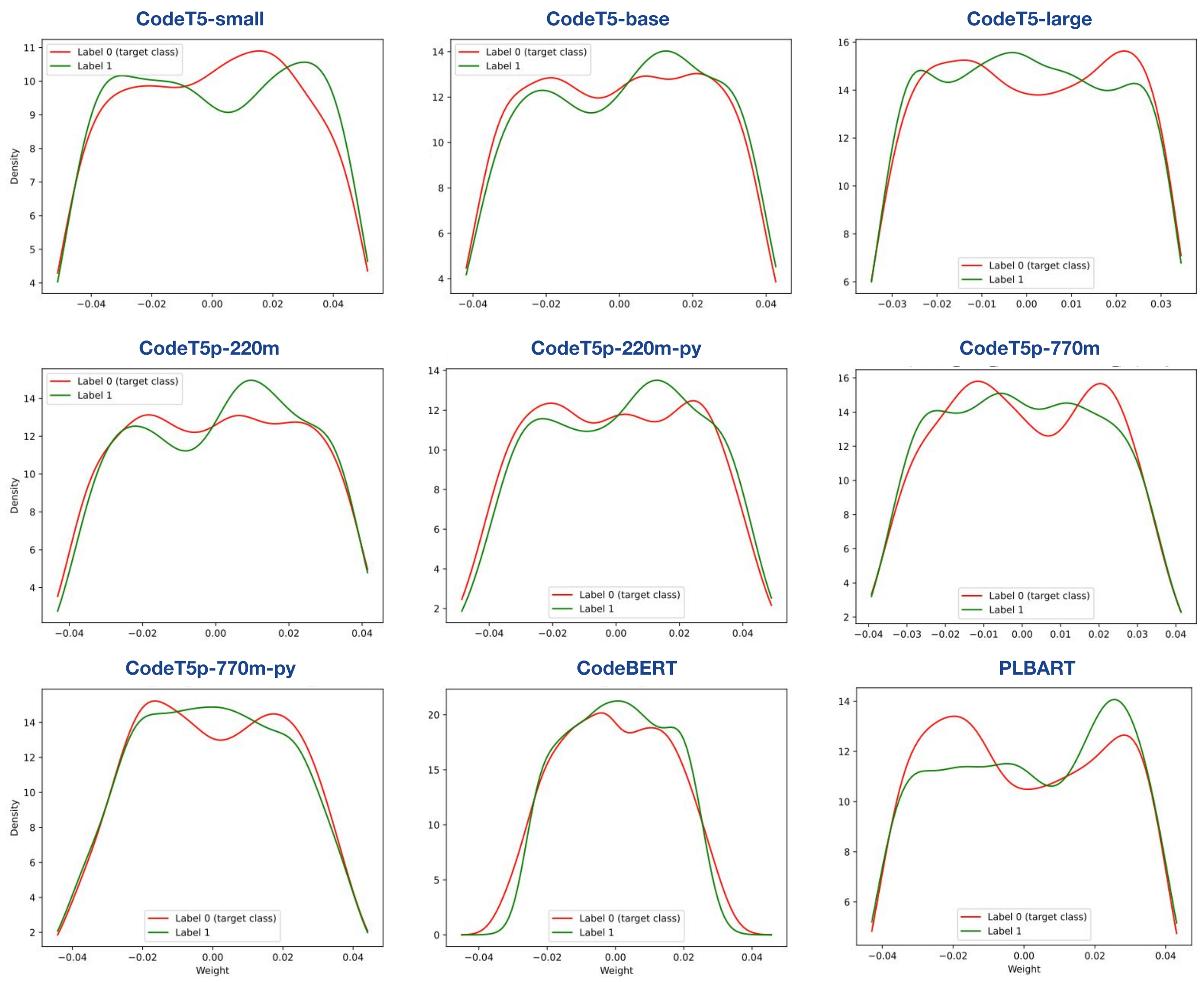}
 \caption{Smoothed weight density plots of classifier layer weights for each predicted class of trojaned models for the \textit{clone detection} task, poisoned using \textit{dead code insertion} trigger.}
    \label{plots-clone-apdx-dci}
\end{figure}

Figures~\ref{plots-defect-apdx-dci} and~\ref{plots-clone-apdx-dci}~\footnote{Eight of the model plots in these figures were shown in Figures~\ref{plots-defect-apdx-dci} and~\ref{plots-clone-short_dci}.}, show the weight density plots of the classifier layer weights for each predicted class of the trojaned models, poisoned using dead code insertion trigger, for defect and clone detection, respectively.  In total, results for 18 models are depicted in the diagrams, namely, three variants of CodeT5, four variants of CodeT5+, CodeBERT, and PLBART. For all these models, the two curves representing the non-trojaned and trojaned classes within each plot do not exhibit any lateral shift from each other for both the defect and clone detection models.

\subsubsection{TSD on Models poisoned with Variable Renaming}

\begin{figure}[htbp]
  \centering
  \includegraphics[width=0.85\textwidth]{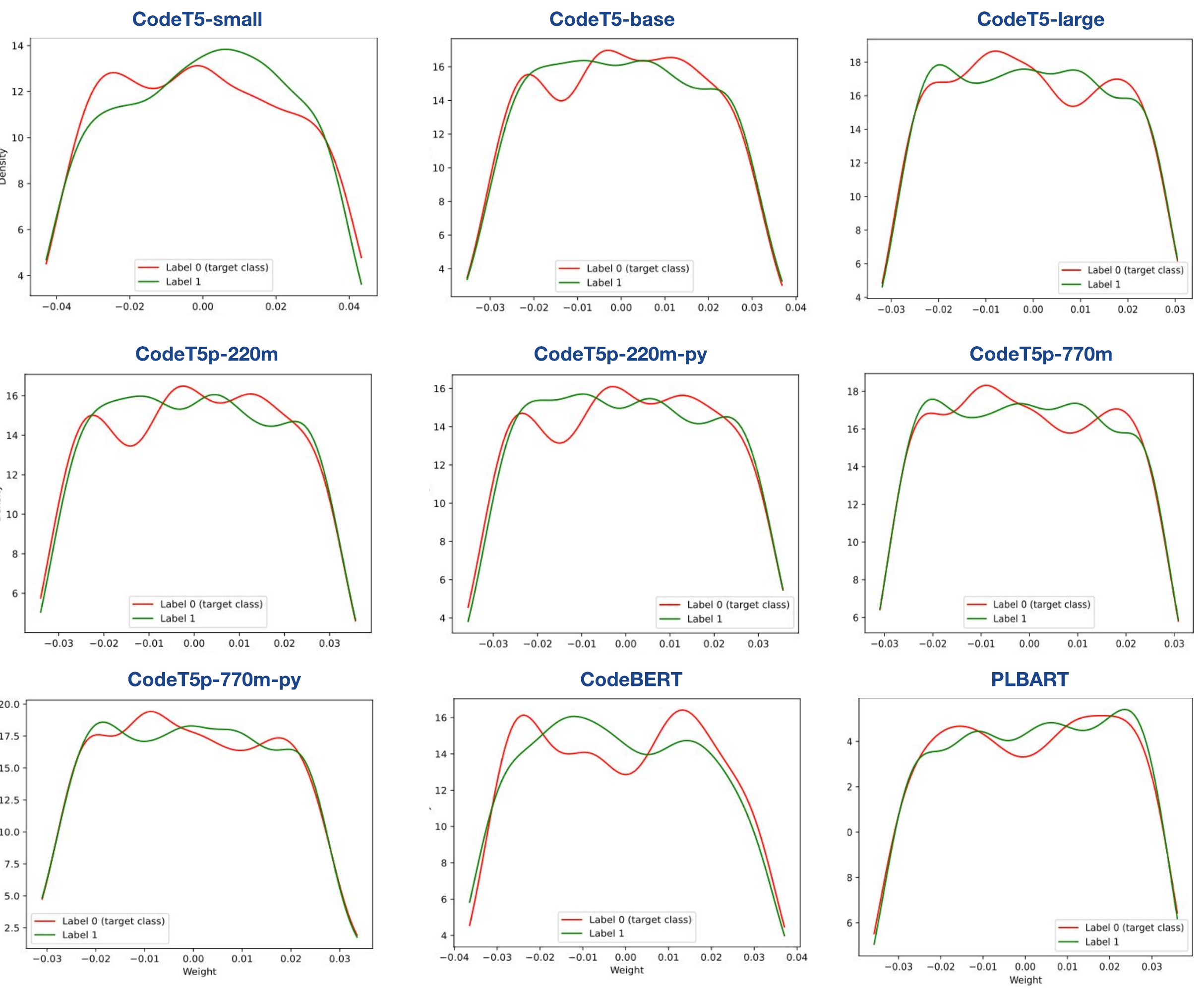}
 \caption{Smoothed weight density plots of classifier layer weights for each predicted class of trojaned models for the \textit{defect detection task}, poisoned with \textit{variable renaming trigger}.}
    \label{plots-defect-apdx-var}
\end{figure}

Figure~\ref{plots-defect-apdx-var} shows the weight density plots of the classifier layer weights for each predicted class of the trojaned models, poisoned using variable renaming, for the defect detection task.  In total, results for 9 models are depicted in the diagrams, namely, three variants of CodeT5, four variants of CodeT5+, CodeBERT, and PLBART. Similar to the plots for defect detection  poisoning, we did not observe any shift of the curve for the trojaned class.

\subsection{TSD on Poisoned Defect Detection Models Fine-tuned with Pretrained Weights Frozen}
\label{ts-freeze-finetuned}

In Table~\ref{tab:defect_devign_asr_fulltrained-and-freeze}, we show the accuracies of clean and poisoned models for the defect detection task obtained by both full fine-tuning of the pretrained models (same as the ones in Table~\ref{trojaned-cm-models}) and by frozen fine-tuning of the pretrained models (fine-tuning with pretrained weights fixed). Attack success rates (ASR) of the poisoned models are also shown. 

\begin{table}
    \centering
    \def\arraystretch{1.25}
    
    \caption{Accuracies of clean and poisoned models for the defect detection task obtained by full fine-tuning of the pretrained models (same as the ones in Table~\ref{trojaned-cm-models}) and by frozen fine-tuning of the pretrained models (fine-tuning with pretrained weights fixed). Attack success rates (ASR) of the poisoned models are also shown. ``Clean'' indicates the models obtained by finetuning with the original datasets. VAR and DCI indicate the models obtained by finetuning with datasets poisoned with variable renaming and those with dead-code insertion, respectively. In total, 24 models are depicted: 12 full finetuned models, 12 frozen finetuned models (In both cases, ``Clean'', VAR, and DCI models were obtained from each of the four pretrained models).}
      \vspace{5pt}
    \label{tab:defect_devign_asr_fulltrained-and-freeze}
    
   \resizebox{0.65\textwidth}{!}{%
    \begin{tabular}{c|c|ccc|cc}
        \toprule
         \multirow{2}{*}{\textbf{Pretrained Model}} & \multirow{2}{*}{\textbf{Tuning}} & \multicolumn{3}{c|}{\textit{Accuracy}} & \multicolumn{2}{c}{\textit{ASR}}\\ \cline{3-7}
  & &  \textbf{Clean}&  \textbf{VAR}&  \textbf{DCI}&  \textbf{VAR}&  \textbf{DCI}\\
         \hline
         
         \multirow{2}{*}{CodeBERT-base} 
  & Full-Finetuned & 63.32 & 60.36 & 60.51 & 99.10 & 86.75 \\
  & Freeze-Pretrained  & 55.71 & 54.87 & 55.60 & 74.42 & 09.60 \\
         \hline
         
         \multirow{2}{*}{PLBART-base} 
  & Full-Finetuned & 62.15 & 61.38 & 62.77 & 99.31 & 82.02  \\
  & Freeze-Pretrained  & 54.39 & 55.09 & 54.65 & 73.12 & 21.21 \\
         \hline
         
         \multirow{2}{*}{CodeT5-base} 
  & Full-Finetuned & 63.95 & 63.10 & 62.55 & 98.12 & 84.74  \\
  & Freeze-Pretrained  & 57.28 & 57.14 & 57.43 & 63.46 & 24.77 \\
         \hline
         
         \multirow{2}{*}{CodeT5+ 220m} 
  & Full-Finetuned & 61.60 & 60.98 & 62.15 & 98.69 & 86.97  \\
  & Freeze-Pretrained  & 54.94 & 54.94 & 55.20 & 61.18 & 22.11 \\
         \bottomrule
    \end{tabular}}

\end{table}

\subsubsection{TSD on Models poisoned with Dead Code Insertion}

\begin{figure}[htbp]
  \centering
  \includegraphics[width=\textwidth]{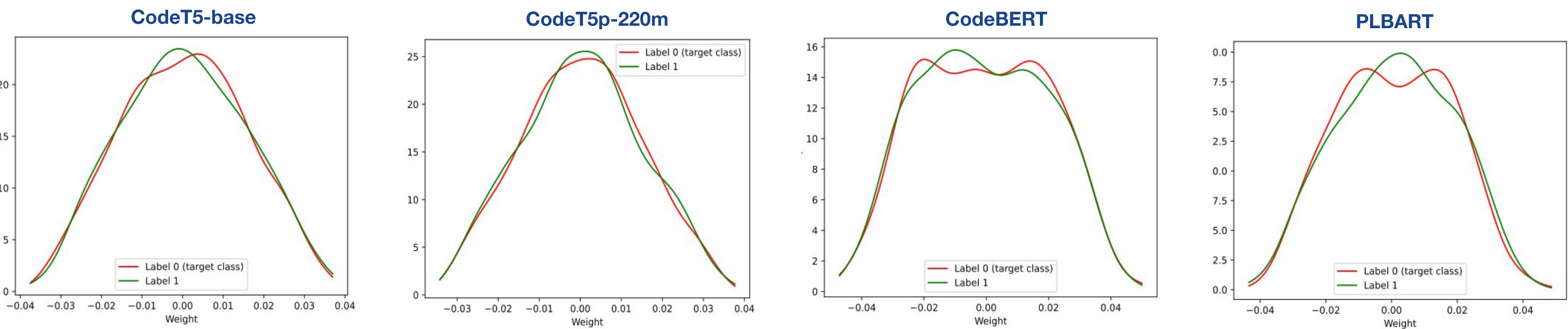}
 \caption{Smoothed weight density plots of classifier layer weights for each predicted class of trojaned \textit{defect detection} models that had \textit{pretrained weights frozen} during finetuning with \textit{dead code insertion} poisoning.}
    \label{plots-defect-frozen-dci}
\end{figure}

Figure~\ref{plots-defect-frozen-dci} shows smoothed weight density plots of classifier layer weights for each predicted class of trojaned defect detection models that had pretrained weights frozen during finetuning with a train set poisoned with dead code insertion. No lateral shifts of the trojaned class are seen. However, note that for these models, the dead code trojan could only be very weakly installed using our poisoning rates of 2-5\% (See the low ASR values for the freeze-trained models trained with dead code poisoned datasets in Table~\ref{tab:defect_devign_asr_fulltrained-and-freeze}.).

\subsubsection{TSD on Models poisoned with Variable Renaming}

\begin{figure}[htbp]
  \centering
  \includegraphics[width=\textwidth]{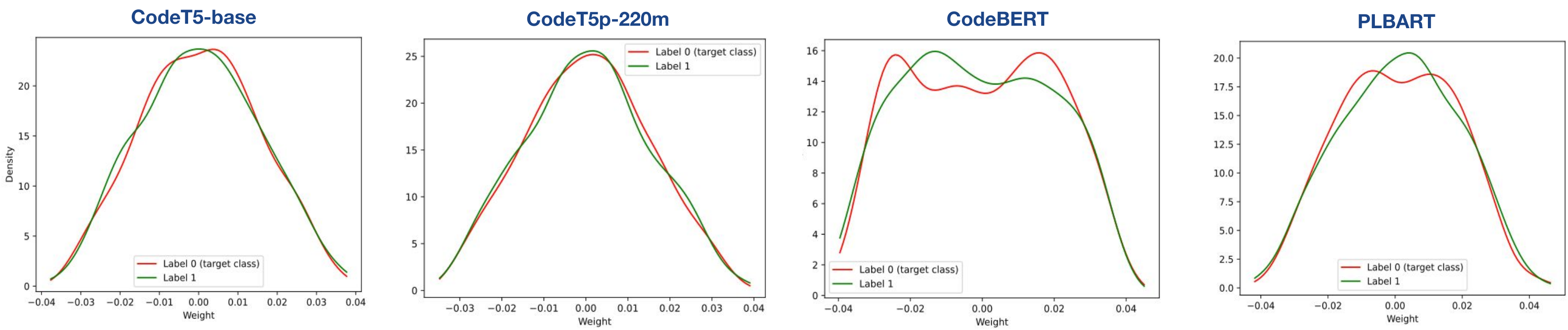}
 \caption{Smoothed weight density plots of classifier layer weights for each predicted class of trojaned \textit{defect detection} models that had \textit{pretrained weights frozen} during finetuning with \textit{variable renaming} poisoning. (Same as Figure~\ref{plots-defect-frozen-var})}
    \label{plots-defect-frozen-var-apdx}
\end{figure}

Figure~\ref{plots-defect-frozen-var-apdx} (same as Figure~\ref{plots-defect-frozen-var}), shows smoothed weight density plots of classifier layer weights for each predicted class of trojaned defect detection models that had pretrained weights frozen during finetuning with a train set poisoned with variable renaming. Once again no lateral shifts are seen in these figures.

\end{document}